\documentclass[aps,prl,twocolumn,superscriptaddress,showpacs]{revtex4-1}

\usepackage{graphicx}
\usepackage{amsmath}
\graphicspath{{./images/}}
\usepackage{color}

\begin{document}

\title{Scattering of bi-flagellate micro-swimmers from surfaces}

\author{Enkeleida Lushi} 
\email{lushi@cims.nyu.edu}
\affiliation{Courant Institute of Mathematical Sciences, New York University, New York, USA}
\author{Vasily Kantsler}
\email{V.Kantsler@warwick.ac.uk}
\affiliation{Department of Physics, University of Warwick, Warwick, UK}
\author{Raymond E. Goldstein}
\email{R.E.Goldstein@damtp.cam.ac.uk}
\affiliation{Department of Applied Mathematics and Theoretical Physics, University of Cambridge, Cambridge, UK}

%\date \today

\begin{abstract}
We use a three-bead-spring model to investigate the dynamics of bi-flagellate micro-swimmers near a surface. While the primary dynamics and scattering are governed by geometric-dependent direct contact, the fluid flows generated by the swimmer locomotion are important in orienting it toward or away from the surface. Flagellar noise and in particular cell spinning about the main axis help a surface-trapped swimmer escape, whereas the time a swimmer spends at the surface depends on the incident angle. The dynamics results from a nuanced interplay of direct collisions, hydrodynamics, noise and the swimmer geometry. We show that to correctly capture the dynamics of a biflagellate swimmer, minimal models need to resolve the shape asymmetry. \\

\end{abstract} 


%\keywords{locomotion, algal surface accumulation, swimming microorganisms}

\maketitle

Microscopic swimming cells such as bacteria, algae and spermatozoa live in porous natural habitats and this geometrical confinement necessitates interactions with surfaces. Surface interactions are fundamental in many biological processes, e.g. biofilm formation and egg fertilization, however despite their ubiquity in nature not all is known about the physics of micro-locomotion in a confined environment \cite{Berke08}. Better understanding of the dynamics of micro-organisms near surfaces is crucial for many applications to control microbial locomotion \cite{Rusconi14}.

There has been considerable discussion on whether Brownian and intrinsic noise, long-range hydrodynamic or short-range mechanical forces determine the surface interactions of these microorganisms \cite{Berke08, Li09, Drescher11, Dunstan12, Spagnolie12, Kantsler13, LiArdekani14, Spagnolie15, Contino15, Goldstein15}. Micro-swimmers such as bacteria or spermatozoa accumulate near boundaries \cite{Berke08, Smith09} and navigate alongside them \cite{Denissenko12, Molaei14, Sipos15}. Hydrodynamic and steric interactions with surfaces are credited for the circling behavior of individual bacteria in surfaces\cite{Lauga06, DiLeonardo11, Lopez14}, the boundary-following of artificial motile colloids \cite{Takagi14, Brown14, Das15} and the emergence of collective motion in confined bacterial suspensions \cite{Wioland13, Lushi14, Wioland16a, Wioland16}. Direct measurements of the flow fields around an individual bacterium reveal that the intrinsic stochasticity of its motion drowns the effects of long-range fluid dynamics and implies that surface interactions of bacteria are dominated by the direct collisions and Brownian noise \cite{Drescher11, Drescher10}. Eukaryotic swimmers such as algae {\it C. reinhardtii} scatter off a surface by pushing against it with their flagella \cite{Kantsler13, Contino15}, indicating that the primary surface dynamics is governed by direct or steric interactions. For {\it C. reinhardtii} it was shown that the swimmer geometry and flagella length are crucial in determining the surface scattering \cite{Kantsler13}.  

\begin{figure}
\centering
\includegraphics[width=0.45\textwidth]{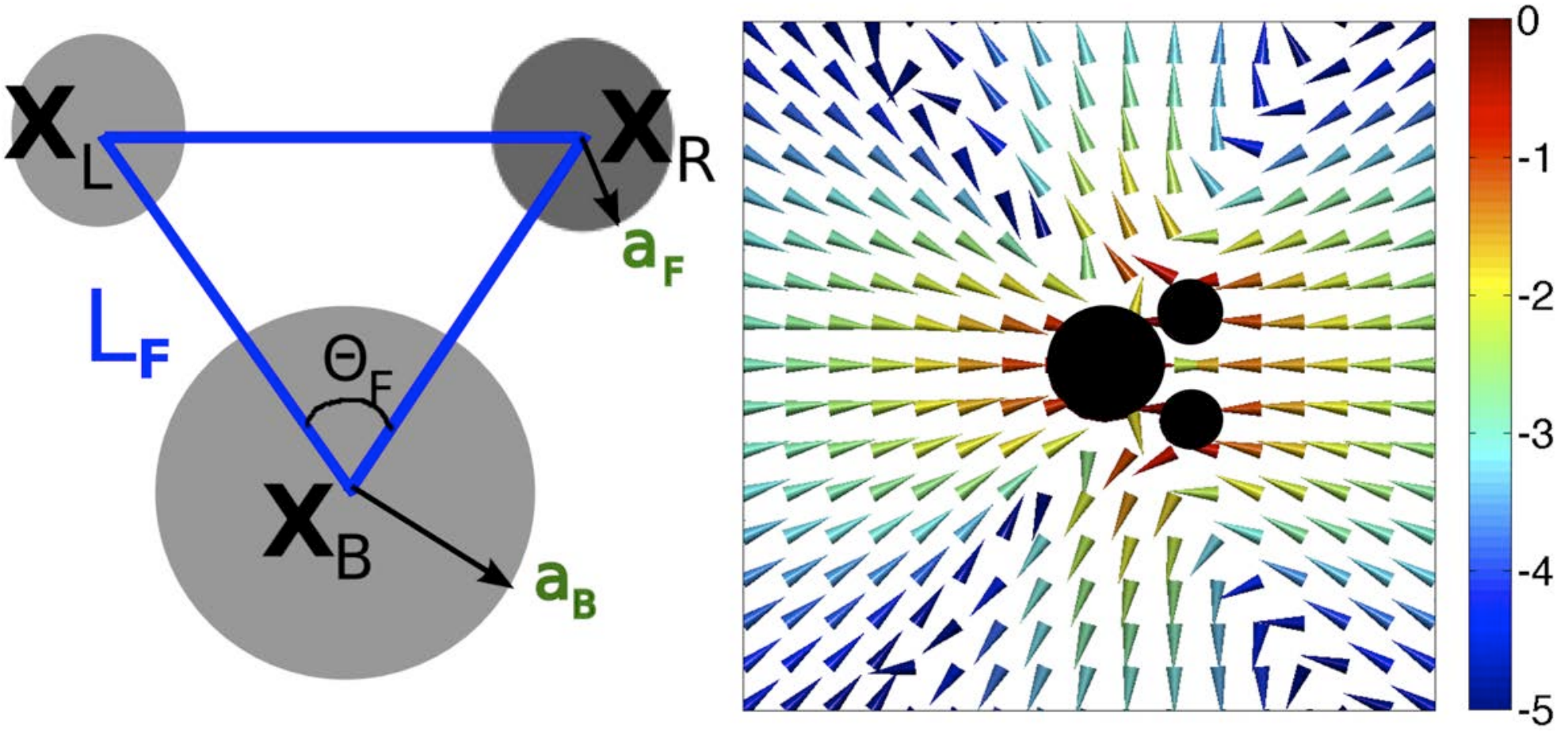}
\vspace{-0.1in}
\caption[width = 0.55\textwidth]{(Color online) Left: diagram of the three-bead swimmer. Right: the disturbance fluid flow in free space; the color represents flow magnitude in a logarithmic scale. }
\vspace{-0.in}
\label{Fig1}
\end{figure}

Hydrodynamics however, while seemingly drowned out by the noise and obscured by the complex surface interaction, is ever present and affects the swimmer motion. Using a simple model biflagellate micro-swimmer, we investigate here the delicate interplay between the swimmer geometry, flagella, the generated fluid flows and noise in its interaction with a surface. The swimmer, consisting of three beads connected by elastic springs (see Fig. \ref{Fig1}), moves with constant propulsive flagellar forces, and generates a ``puller'' disturbance fluid flow resembling the experimental observations \cite{Drescher10, Guasto10}. We show that while the primary scattering dynamics is indeed governed by the swimmer geometry and steric effects, the attraction or repulsion of the swimmer to surfaces is influenced by fluid dynamics. The scattering angle is shown to depend on the swimmer geometry, as also observed in experiments \cite{Kantsler13}. We show that noise and in particular cell spinning help surface-trapped swimmers escape. The time that a puller biflagellate swimmer spends at a surface is shown to depend quasi-linearly to the incoming angle, a result confirmed with new experimental measurements. 

We also establish here the appropriate levels of simplicity in micro-swimmers models so that they capture the behavior of bi-flagellates as it observed in experiments with {\it C. reinhardtii}. Since the dynamics near a wall depends primarily on direct contact of the flagella, we show that an asymmetric or triangular shape is needed to capture the scattering phenomenon. Moreover, since such swimmers spin about their axis and scattering depends on the cell configuration \cite{Contino15}, we show that three-bead models are the most minimal that can still display the correct physical behavior. 

{\bf Model.--}  We consider a minimal three-bead-and-springs model for {\it C. reinhardtii} as illustrated in Fig. \ref{Fig1}A. The model is inspired by the recent experimental measurements of the flow field around this type of algae \cite{Drescher10, Guasto10} which can be approximated reasonably well by
three Stokeslets. The three-Stokeslet description has been recently used to investigate phenomena such as the coupled flagella dynamics and synchronization \cite{Friedrich12, Leptos13, Wan14}, tumbling \cite{Bennett13} and phototaxis \cite{Bennett15, Jibuti14} in {\it C. reinhardtii}.

We assume constant propulsive forces are concentrated on the flagella beads with radius half the body-bead. Using a zero force condition on the entire swimmer and a balance of forces on each bead (similar to the two-bead puller model of Hernandez-Ortiz {\it et al.}\cite{HOrtiz05,HOrtiz09}), we can derive the equation of motion for each bead. The coupled dynamics of the beads $B,L,R$ (denoting the swimmer body, left and right flagellum beads respectively) can be compactly described as follows
\begin{align}\label{xdot}
\frac{d \mathbf{x}_k}{dt} &=  \frac{1}{\xi_k} \left[ \delta_{k, (L,R)} \mathbf{f}^f_k + \mathbf{f}^c_k + \mathbf{f}^x_k  \right] \nonumber\\
                                       &+ \sum_{j=(B,L,R)} G_{a_j} (\mathbf{x}_k, \mathbf{x}_j) \left[ (1- \delta_{k,j}) \mathbf{f}^c_j + \mathbf{f}^x_j   \right] \nonumber \\
                                       &+ \sum_{j=(B,L,R)} \tilde{G}_{a_j} (\mathbf{x}_k, \mathbf{x}_j) \left( \mathbf{f}^c_j+ \mathbf{f}^x_j   \right)  %       &+ \text{flow due to images if near a wall}
\end{align}
for $k=(B,L,R)$. Here $\xi_k = 6 \pi a_k$ are Stokes drag coefficients and $a_B=1/3$, $a_L=a_R:=a_F=1/6$ are the bead radii. The propulsive forces $\mathbf{f}^f_L$ and $\mathbf{f}^f_R$  act only on the two flagella beads $L$ and $R$. The connector spring forces $\mathbf{f}^c$ are calculated using the finitely extensible nonlinear elastic (FENE) spring model introduced by Hernandez-Ortiz {\it et. al} in \cite{HOrtiz09} for their two-bead-spring swimmers. The spring constant is $h=25$ and the springs are finitely-extensible with $L_{max}/L=1.01$ and $L_{min}/L=0.99$. The steric forces $\mathbf{f}^x_k$ are calculated with the short-ranged and purely repulsive Weeks-Chandler-Anderson potential \cite{WCA} activated at $2^{1/6}a \approx 1.12a$ distance from a bead's center to prevent overlaps. Other potentials can be used as well, but they should not be long ranged.
In Eq. (\ref{xdot})
\begin{align}%\label{stokeslet}
G_{a} (\mathbf{x}_k, \mathbf{x}_j) = 
\frac{1}{8\pi} \left[ \frac{r^2 + 2a^2}{(r^2 + a^2)^{3/2}}\textbf{I} 
+\frac{(\textbf{x}_k-\textbf{x}_j)(\textbf{x}_k-\textbf{x}_j)^T}{(r^2 + a^2)^{3/2}} \right] \nonumber
\end{align}
with $r=||\textbf{x}_k-\textbf{x}_j||$ is a regularized Stokeslet in 3D where the regularization parameter $a$ is the bead radius \cite{Cortez05}. If the swimmer is near a no-slip wall, the method of images with regularized image Stokeslets  \cite{Ainley08}, here denoted by $\tilde{G}_{a}$, are employed. The fluid velocity at some point $\mathbf{x}_e$ is
\begin{align}\label{fluidvel}
 \mathbf{u}(\mathbf{x}_e) &= \sum_{j=(B,L,R)} [ G_{a_j} (\mathbf{x}_e, \mathbf{x}_j)+\tilde{G}_{a_j} 
                                          (\mathbf{x}_e, \mathbf{x}_j)   ]\left( \mathbf{f}^c_j + \mathbf{f}^x_j   \right).\nonumber 
 \end{align}
If steric forces, necessary in problems involving confinement, are included in the model, then the  Stokeslets need not be regularized. However, for fast simulations of many such swimmers in the bulk, the regularized Stokeslet is an easy way to avoid there costly computations of steric interactions while still retaining some information about the swimmer size. We include them in results shown here.
 
The wall is at $z=0$ and the swimmer in the $z>0$ half-space. The dynamics is in 3D, but, in the absence of body and flagella rotation, the swimmer motion remains planar and with direction $ \mathbf{n}= (\mathbf{x}_L+\mathbf{x}_R)/2-\mathbf{x}_B $ for a puller swimmer. Propulsive flagella forces $\mathbf{f}^f_L$ and $\mathbf{f}^f_R$ (cf Fig. \ref{Fig1}A) are taken to be $-\mathbf{n}/|\mathbf{n}|/2$ and fixed in magnitude. 

For a ``pusher'' swimmer, like the moving-backward-only mutant {\it Chlamydomonas} CC-2679 mbo1 \cite{Kantsler13}, we reverse the direction of the applied flagellar forces $\mathbf{f}^f$ (see Fig. \ref{Fig1}A). The swimmer then moves head-bead first with direction $ \mathbf{n}= \mathbf{x}_B -(\mathbf{x}_L+\mathbf{x}_R)/2 $. The generated flow field generated then is ``pusher''-like and with the direction reversed from the puller case in Fig. \ref{Fig1}B.

{\bf Materials and Methods.--} The experiments' protocol followed here is exactly that followed in the forerunning study of Kantsler {\it et al.} \cite{Kantsler13}. {\it C. reinhardtii} strains CC-125 WT, CC-2347 shf1-277, CC-2289 lf3-2, and CC-2679 mbo1 (The Chlamydomonas Resource Center, {\it www.chlamy.org}) were grown and used. Quasi-2D microfluidic channels were manufactured with standard soft lithography techniques. Swimming characteristics of individual algae cells and their trajectories were reconstructed by applying a custom-made particle tracking-velocimetry algorithm to image data taken with a Nikon TE2000-U inverted microscope (10? objective, 10 fps). The flagella dynamics close to the boundary were captured with a Fastcam SA-3 Photron camera (500�2,000 fps, 40$\times$/NA 1.3 oil immersion and 60$\times$/NA 1.0 water immersion objectives).

\begin{figure*}[] 
\centering
\includegraphics[width=2\columnwidth]{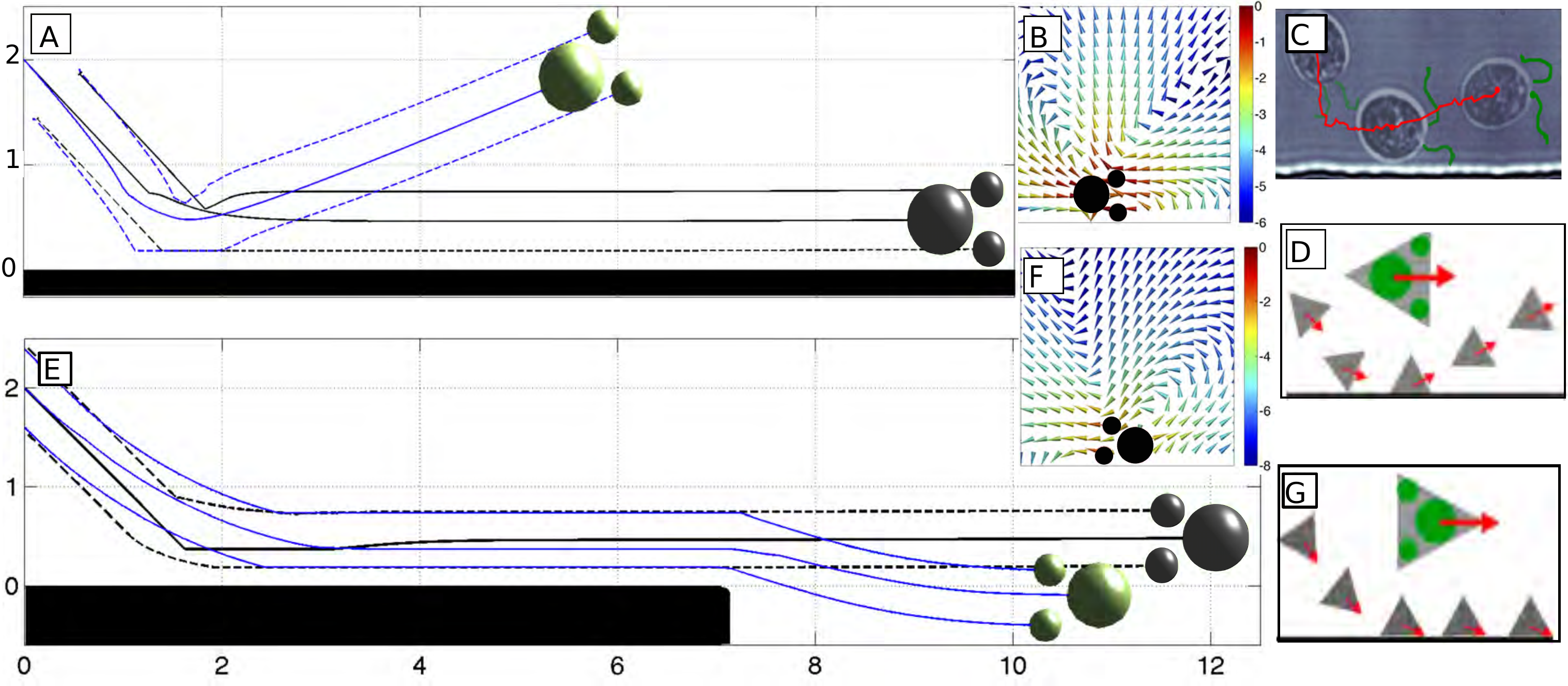}
\vspace{-0.0in}
\caption{  (Color online)  (A) Trajectories of a puller swimmer with and without hydrodynamics (beads green/gray and their trajectories blue/black respectively). (B)  Fluid flow generated when the puller swimmer hits the wall. (C) Scattering trajectory of a wild-type {\it C. reinhardtii} \cite{Kantsler13}. (E) Trajectories of a pusher swimmer with and without hydrodynamics (beads in green/gray and their trajectories in blue/black respectively), along the wall and past a smoothed $90^o$ corner. (F) Fluid flow generated by the pusher swimmer at the wall. (D,G) The swimmer triangular geometry scattering argument suggested by Kantsler {\it et.al.} \cite{Kantsler13}.  See the Supplementary Material \cite{Supplementary} for movies of the dynamics.}
\vspace{-0.0in}
\label{Fig2}
\end{figure*}

\begin{figure*}[]
\centering
\includegraphics[width=2\columnwidth, height=3.3in]{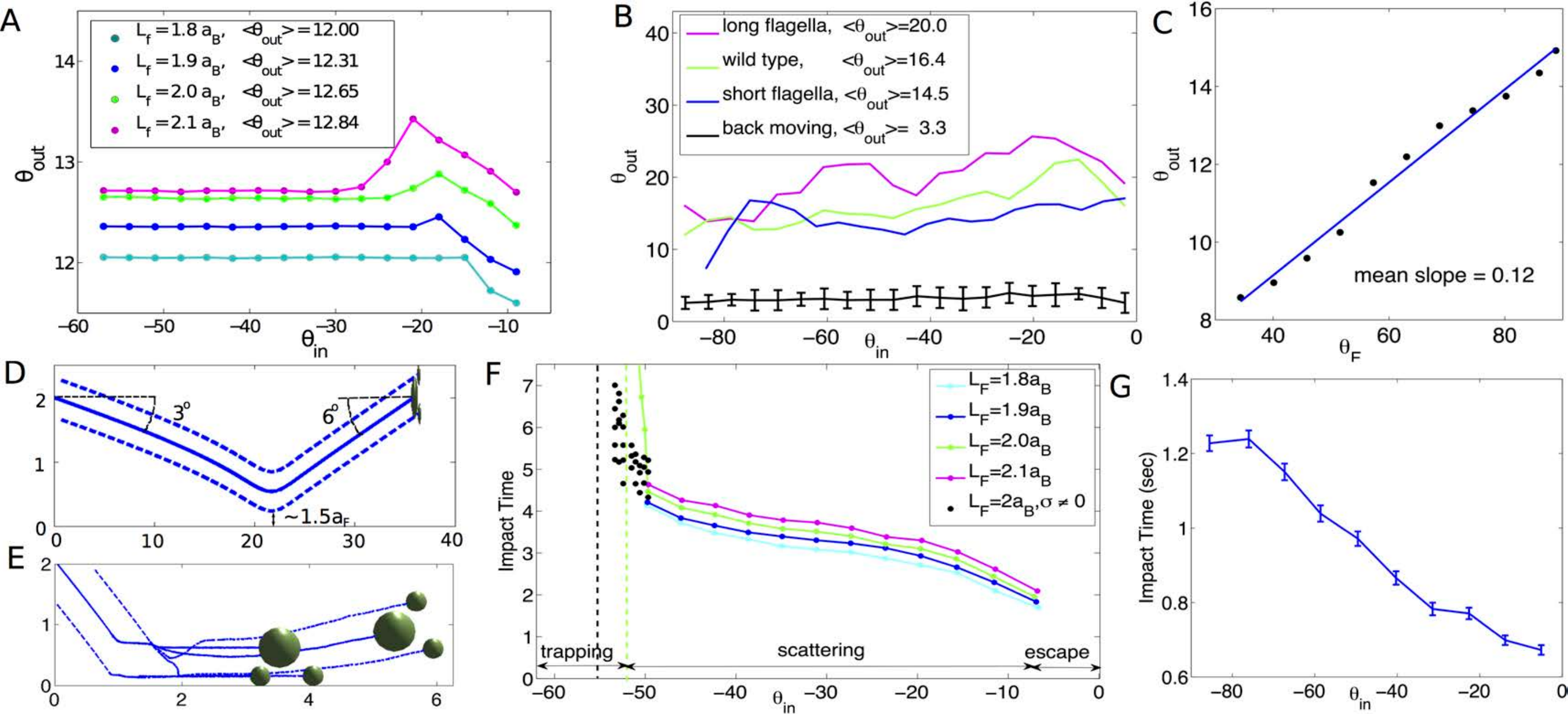}
\vspace{-0.0in}
\caption{  (Color online)  (A) Scattering angles of a puller swimmer vs. incident angles for four different swimmer flagella length and fixed inter-flagella angle $\theta_F=70^o$. (B) Experimentally measured scattering angles vs. incident angles for four types of {\it C. reinhardtii}: three puller types with different flagella lengths and one pusher type. (C) Scattering angle for a variety of inter-flagella angle $\theta_F=70^o$ keeping fixed the incidence angle $\theta_{in}=45^o$ and flagella length $L_F=2a_B$. (D) A swimmer hydrodynamically interacting with the wall without touching it. (E) A swimmer is trapped at the wall, but it can scatter if noise is included in the dynamics. (F) The impact time of a puller swimmer at the wall vs. incident angles for different flagella lengths and with noise. (G) Experimentally measured impact time for wild-type {\it C. reinhardtii}.  Angles are reported in degrees.
}
\vspace{-0.in}
\label{Fig3}
\end{figure*}

{\bf Fluid Dynamics.--}  Fig. \ref{Fig2} illustrates the swimmer dynamics near a wall. Pusher and puller swimmers behave differently: puller swimmers scatter off the wall as in Fig. \ref{Fig2}A which resembles experimental observations Fig. \ref{Fig2}B, whereas pushers tend to swim along-side the wall, as in Fig. \ref{Fig2}D. Surprisingly, this behavior partly depends on the generated fluid flows. If the beads interact only through the springs and the hydrodynamics is neglected, the puller swimmer remains at the wall and does not scatter, Fig. \ref{Fig2}A. Without hydrodynamics, a pusher swimmer does not notice the presence of a corner in Fig. \ref{Fig2}D. In experiments, a spermatozoon, which is a pusher, is observed to swim past a corner making a negative angle to the horizontal \cite{Kantsler13}. Our model captures this phenomenon, as seen in Fig. \ref{Fig2}D, and demonstrates that it results from the hydrodynamic attraction to the vertical side of the corner The corner is rounded with a radius similar to the flagella beads for ease of computation and interactions are computed accordingly. The importance of fluid dynamics is obvious in the swimmer orientation as it approaches the boundary: the pusher in Fig. \ref{Fig2}D tends to align parallel with the wall, whereas the puller in Fig. \ref{Fig2}A tends to orient perpendicular to it.

{\bf Scattering.--}  In Fig. \ref{Fig2}A we show a typical trajectory of a three-bead puller swimmer when the incoming angle $\theta_{in}$ before the collision is not steep ($\theta_{in}+\theta_F/2 < \pi/2$). At the wall the swimmer executes a turn. The lower flagellum bead experiences more drag (as seen by the asymmetry of the configuration in Fig. \ref{Fig3}BC), and slides along the wall. Spring forces push on the other flagellum and body bead seeking to return to the rest configuration. This results in the swimmer turning at the wall in a finite time and, once re-oriented, swimming away from the wall.

We measure the scattering angle $\theta_{out}$ of a for a variety of incidence angles $\theta_{in}$ and swimmer flagella lengths $L_F$ but keeping fixed the inter-flagella angle $\theta_F=70^o$. The results are shown in Fig. \ref{Fig3}A. Angles are measured at a distance $1.2a_F$ above the wall and are with respect to the horizontal surface, while ``touching'' of the wall is implied when the steric forces are activated. (Changing the criterion for measuring the incoming and outgoing angles changes the results only qualitatively and minimally.)  We notice in in Fig. \ref{Fig3}A that the angle at which a swimmer scatters of a wall does not seem to depend on the incidence angle, with the angles differing by less than $5$ degrees. This suggests that the memory of the incoming angle is lost during the swimmer turning at the wall, especially for steep incident angles.% when the time spent re-orienting at the wall is longer.%, as indicated by the essentially flat part of $\theta_{out}(\theta_{in})$ in Fig. \ref{Fig2}A. 

For completeness, we plot in Fig. \ref{Fig3}B the scattering angle vs. the incident angle for four different strains of {\it Chlamydomonas}: the WT (wild type) CC-125 with flagella length $11-13 \mu m$, the short-flagella mutant CC- 2347 shf1 with flagella length $6-8 \mu m$, the long-flagella mutant CC-2289 lf3-2 with flagella length $12-22 \mu m$,  and the moving-backward-only mutant CC-2679 mbo1 which is considered a ``pusher''-swimmer just like bacteria \cite{Kantsler13}. The trajectory angles are measured at a distance $20\mu m$ from the wall. In the experiments the scattering angle of the puller-swimmer does not differ much with the incident angle, but increases with the flagella length. The scattering angle for the pusher-swimmer is very low ($~5^o$), indicating they swim along the wall just like bacteria.

The scattering angle in the model here does not vary much with the flagella length, as seen in Fig. \ref{Fig3}A, whereas in the experiments the scattering angle  increases with the flagella length, as seen in Fig. \ref{Fig3}B and analyzed in \cite{Kantsler13}. Since the flagella in {\it C. reinhardtii} are not straight but are flexible, their change in length is also a change in the swimmer triangular geometry. With the model we can investigate the importance of the geometric shape and direct ciliary contact with the wall by varying the inter-flagella angle $\theta_F$ but fixing the flagella length $L_B=4a$ and incidence angle $\theta_{in}=45^o$. Fig. \ref{Fig3}C shows a pronounced increase of the scattering angle with increasing $\theta_F$ confirming the argument that scattering depends mostly on the swimmer triangular geometry just as proposed by Kantsler {\it et al.} \cite{Kantsler13} and illustrated in Fig. \ref{Fig2}FG. 

%{\it C. reinhardtii}'s flagella are flexible and execute a periodic stroke while the three-bead model here does not account for that. 

{\bf Impact Time.--} We discuss the time it takes a puller swimmer to turn at the wall, or impact time, which is measured as the time during which a flagellum touches the wall. We observe three distinct states: (i) Hydrodynamic attraction of the swimmer to the wall when starting with quasi-horizontal configuration, as illustrated in Fig. \ref{Fig3}D and then escape from the wall without touching it. The angle of escape is different from the angle of approach, indicating the effect of the asymmetrical placement of the Stokeslets. The impact time in this regime is exactly zero. (ii) An intermediate regime of scattering dynamics where the dependence of the impact (turning) time to the incident angle almost linear; the swimmers approaching with a steep angle take more time to turn at the wall. Fig. \ref{Fig3}F shows this impact time for a variety of incident angles and flagella lengths but fixed inter-flagella angle $\theta_F=70^o$. (iii) A regime where the swimmer gets trapped at the surface with the flagella touching the wall, as seen in Fig. \ref{Fig1}E. This typically happens for $\theta_{in}>(\pi-\theta_F)/2$. The impact time is infinite, thus there is a vertical asymptote, as shown in Fig. \ref{Fig3}F. 

We measured the impact time for wild-type (WT) {\it Chlamydomonas} for a variety of incident angles, as shown in Fig. \ref{Fig3}G. If the incident angles not very steep, the impact time also grows quasi-linearly. However, the impact time tapers off for high incident angles, indicating that the swimmers in the experiments do not become trapped.% at the surface like the model-swimmers do. 

{\bf Noise.--}  
It has been shown that in {\it C. reinhardtii} the flagella pair apparent asynchronization and slips result in randomization of the micro-organism motion \cite{Polin09}. Flagella proximity to a surface might also significantly increase the pair asynchronization probability. We can incorporate this intrinsic noise in our model by including noise with strength $\sigma$ in the flagella bead dynamics as $d \mathbf{x}_k/dt = (d \mathbf{x}_k/dt )_{deterministic}+ \xi_k(t)$ where $\xi_k(t)$ has a Gaussian probability distribution with zero mean and correlation function $<\xi_k(t),\xi_j(t')>\sigma^2\delta_{kj}\delta(t-t')$. 

We illustrate how sufficient noise can help a swimmer overcome surface trapping. In Fig. \ref{Fig3}E we show an example where noise added to the flagellar dynamics enabled escape whereas the same initial condition for a noiseless swimmer results in it being bound to the surface. Note that the position of the vertical asymptote in the impact time vs. incident angle plot shown in Fig. \ref{Fig3}F is pushed further to the left for examples with flagellar noise, illustrating how sufficient noise can help in the marginal cases. For {\it C. reinhardtii} however there are other robust mechanisms that can help swimmers free themselves from surface entrapments \cite{Contino15}, which we discuss later.

\begin{figure}[]
\centering
\includegraphics[width=0.46\textwidth]{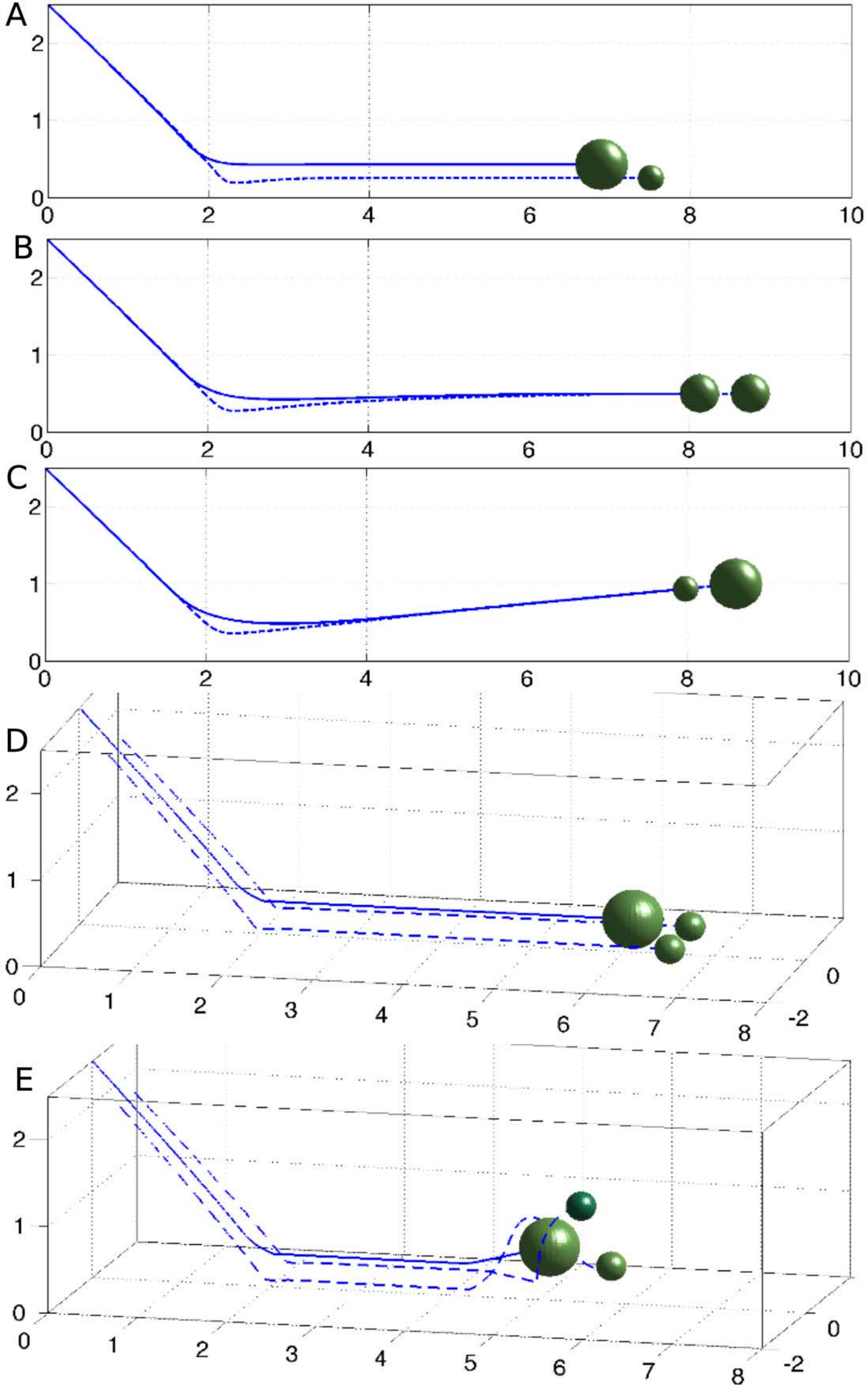}
\vspace{-0.1in}
\caption{  (Color online) Micro-swimmer dynamics near a wall: comparisons of trajectories obtained with different models. (A) Asymmetric ``big-tail'' two-bead-and-springs puller swimmer with body-bead twice the radius of the flagellum bead. (B) Symmetric two-bead-and-springs puller swimmer where the body and flagellum beads have same radii. (C) ``Big-head'' two-bead-and-springs puller swimmer with body-bead half the radius of the flagellum bead. (D) Three-bead-and-springs puller swimmer as in Fig. \ref{Fig1} where the swimmer flagella configuration is parallel to the surface gets trapped there. (E) Cell spinning enables the swimmer to escape the wall. See the Supplementary Material \cite{Supplementary} for movies of the dynamics in case E and also another case of spinning-enabled escape.
}
\vspace{-0.1in}
\label{Fig4}
\end{figure}

{\bf One, two or three beads?--} We compare the dynamics and trajectories of bead-and-spring puller swimmers near a wall. The two-bead model has been used to study the collective behavior of front-back symmetric swimmers \cite{HOrtiz05,HOrtiz09}, i.e. the swimmer body and flagellum beads are of equal size. A symmetric two-bead puller, shown in Fig. \ref{Fig4}B, reaches the wall but due to hydrodynamical interaction with its own image, repels away from it while still oriented quasi-parallel to the wall. 

A ``big-tail'' swimmer where the front flagellum bead is smaller than the back body bead, collides with the wall and moves along it, as in Fig. \ref{Fig4}A. It cannot turn away from the wall due to its shape asymmetry and hence is trapped. A ``big-head'' swimmer where the front flagellum bead radius is bigger, broadly mimics the triangular geometry of {\it C. reinhardtii} \cite{Wensink14}. Such a puller swimmer (with hydrodynamics accounted for) can scatter away from the wall. The escape angle is determined by its geometry, as seen in Fig. \ref{Fig4}C. This confirms that an asymmetric swimmer shape is indeed needed to obtain scattering dynamics.

 ``Big-head'' swimmers however are axisymmetric and do not adequately model the geometry of a biflagellate like {\it C. reinhardtii} whose generated fluid flows are better captured by three Stokeslets \cite{Drescher10, Guasto10}. While both the ``big-head'' model and the three-bead model capture the scattering dynamics, the latter captures the dynamics of both flagella which clearly do not remain symmetric when touching a wall.

 {\bf Cell Configuration and Spinning.--}In the cases presented in Fig. \ref{Fig2}, the plane of motion for the triangular swimmer is perpendicular to the wall plane. If the swimmer plane of motion is instead so that flagellar plane is parallel to the wall (see Fig. \ref{Fig4}D), the swimmer has two possible dynamics scenarios: (a) if the incoming angle is steep, them it gets trapped at the wall and does not scatter, or (b) if the incoming angle is quasi-parallel to the wall, then it experiences a slight attraction at the wall due to hydrodynamics, but never touches it. The trapping  case is conceptually similar to that of Fig. \ref{Fig4}A where the swimmer cannot turn away from the wall, whereas the hydrodynamic attraction case is similar to that of Fig. \ref{Fig3}E. Note that in this configuration the swimmer does not have a scattering phase at all! Since {\it C. reinhardtii} has not been observed in experiments to get trapped at the wall (see Fig. \ref{Fig3}G), the trajectory in Fig. \ref{Fig4}D indicates that other mechanisms may prove important in resolving the dynamics at the wall \cite{Contino15, Jeanneret16}.

Comparing the simulations and experiments in Fig. \ref{Fig3}FG, we notice {\it C. reinhardtii} spends a finite time at the wall and eventually escapes, whereas the model swimmer can get trapped there indefinitely. The discrepancies may result from the fact that  {\it Chlamydomonas} slowly spins about its main axis and it was recently observed in experiments that this feature enables the swimmer to escape cylindrical obstacles with diameter about $25 \mu m$ \cite{Contino15}. 

We can check the cell spinning effect with our model here by adding a slow rotation of the flagellar beads about the swimmer's main axis of motion, akin to the effective torque discussed by \cite{Schaar14} to mimic flagellar activity. We test the spinning three-bead-and-spring model on the case of the trapped swimmer of Fig. \ref{Fig4}D and turn on the rotation after the swimmer has spent some time moving alongside the wall. After the cell has rotated by about $90^o$, it is effectively in the configuration where he flagellar plane is perpendicular to the wall, which as discussed  before and illustrated in Fig. \ref{Fig2}ACE, allows the swimmer to scatter of the wall due to its triangular shape. When the swimmer in the the trapped position shown in Fig. \ref{Fig3}E, cell spinning allows it to first get in the position shown in \ref{Fig4}D, and then scatter similarly to Fig. \ref{Fig4}E.  The movies in the Supplementary Material \cite{Supplementary} illustrate this effect: no matter the initial swimmer configuration, if the swimmer is ``trapped'' at the wall, cell-spinning allows it to turn to a favorable position and scatter. Cell spinning thus enables bi-flagellate swimmers to escape from surfaces.

{\bf Discussion.--}
We introduced a model for micro-swimmers where the body and two flagella are represented by spheres connected by elastic springs. The dynamics of a  biflagellate swimmers near straight surfaces was investigated and the roles of the direct collisions, fluid flows and noise are discussed in detail. Some recent studies \cite{Wensink14, Wysocki15} consider the swimmer shape asymmetry, e.g. ``big-head'' swimmers, but they do not include hydrodynamics. Others consider asymmetric hydrodynamics, but neglect asymmetric shape interactions \cite{Tsang15, Tsang16}. Here we show that the dynamics of biflagellates near surfaces is a complex interplay of all these ingredients (shape, hydrodynamics, noise), and the roles of each ingredient should not be neglected in models and computations involving swimmers near surfaces. In particular, our results show that  the swimmer size and shape should be resolved to obtain the correct dynamics near surfaces. More simplified micro-swimmer models, such as a one point-dipole or a squirmer swimmers may be inadequate.

Just as recent experiments suggest \cite{Kantsler13, Contino15}, we find that the scattering of puller swimmers like {\it C. reinhardtii} depends primarily on its triangular geometry and the direct flagellar contact with the surface, and secondarily on the hydrodynamic interaction with the surface. While in the {\it C. reinhardtii} swimmer the scattering angle depends on the flagella length, in the model three-bead-and-spring swimmer it depends on the inter-flagella angle. Both these show that the triangular swimmer geometry is an important factor in determining the scattering angle. Noise is shown to help model swimmers escape surface entrapment, however it is the cell spinning about its main axis that enables the swimmer to escape from the wall. The time a puller swimmer spends to turn at the wall is shown to depend quasi-linearly on the incoming angle, an observation confirmed by new experiments with wild-type {\it C. reinhardtii}. This turning or detention time \cite{Schaar14} at the wall is non-zero and non-constant, indicating that ``point-billiard'' models of micro-organisms just reflecting at the wall are oversimplified and may miss this crucial dynamics. The ability to avoid long-term trapping at surfaces represents a significant advantage for a soil alga like Chlamydomonas, which in its natural habitat navigates a porous material \cite{Contino15}.

The three-bead-and-springs model presented here is versatile and can be adapted to study interactions of bi-flagellates with more complex surfaces and a variety of confinements, e.g. circular chambers as in Ostapenko {\it et al.} \cite{Ostapenko16}, by changing or approximating the form of the Stokeslets in Eq. (\ref{xdot}) to the appropriate ones for the surfaces in question. A time-dependent breast-stroke stroke or a swimmer spinning about its main axis, as observed in the motion of the {\it C. reinhardtii} algae, can be incorporated to model the motion. Moreover, it is possible to include lubrication hydrodynamics to further clarify the motion of the spheric body near a surface \cite{Contino15}. Last, this three-bead-and-springs model can be used to study the collective motion of many bi-flagellates in diverse geometrical confinements and heterogeneous porous media mimicking their natural habitats.

{\bf Acknowledgements.--} We thank M. Contino, J. Dunkel, K. Leptos, M. Polin, and I. Tuval for many helpful discussions. This work was supported in part by an Established Career Fellowship from the Engineering and Physical Sciences Research Council (REG).

\end{document}